\documentclass[a4paper,11pt]{article}

\usepackage{jcappub}
\usepackage[utf8]{inputenc}

\usepackage{bm}

\title{Derivative couplings in massive bigravity}

\author[a]{Xian Gao}

\author[b]{and Lavinia Heisenberg}

\affiliation[a]{Department of Physics, Tokyo Institute of Technology,\\
	2-12-1 Ookayama, Meguro, Tokyo 152-8551, Japan}

\affiliation[b]{Institute for Theoretical Studies, ETH Zurich,\\
	Clausiusstrasse 47, 8092 Zurich, Switzerland
}

\emailAdd{gao@th.phys.titech.ac.jp}
\emailAdd{lavinia.heisenberg@eth-its.ethz.ch}

\abstract{
		In this work we study the cosmological perturbations in massive bigravity in the presence of non-minimal derivative couplings. For this purpose we consider a specific subclass of Horndeski scalar-tensor interactions that live on the unique composite effective metric. For the viability of the model both metrics have to be dynamical. Nevertheless, the number of allowed kinetic terms is crucial. We adapt to the restriction of having one single kinetic term. After deriving the full set of equations of motion for flat Friedmann-Lemaitre-Robertson-Walker background, we study linear perturbations on top of it. We show explicitly that only four tensor, two vector and two scalar degrees of freedom propagate, one of which being the Horndeski scalar, while the Boulware-Deser ghost can be integrated out. 
		}

\keywords{}

\arxivnumber{}

\begin{document}
\maketitle

% % % % % % % % % %
\section{Introduction}

Cosmology became an empirical scientific discipline thanks to the high precision achieved by the cosmological observations. Owing to the increasing number of accurate cosmological data we are in a position to test fundamental physics. Based on the two fundamental pillars, the Cosmological Principle and General Relativity, we established the concordance standard model of the universe to be $\Lambda$CDM, which requires the existence of cold dark matter together with dark energy in form of a cosmological constant. The model serves as an effective cosmological model providing the best fit to current cosmological data \cite{Ade:2015xua,Ade:2015lrj,Amendola:2012ys}. In spite of this excellent agreement the  discovery of some reported anomalies like the lack of power in the region of low multipoles in the Cosmic Microwave Background, the hemispherical asymmetry, the detection of large scale bulk flows and the unexpected large scale correlations might signal the failure of our fundamental assumptions \cite{Ade:2015hxq}. On the other hand, the observed unnatural smallness of the cosmological constant is troublesome to reconcile with standard techniques of quantum field theory \cite{Weinberg:1988cp}. Its instability under quantum corrections renders the model technically unnatural. 

The above mentioned observational and theoretical drawbacks has motivated attempts to explain the late time cosmic acceleration by resorting to dynamical fields or modified gravity.
Modifications in form of an additional scalar field are the most extensively explored ones, since the construction of interactions is less burdensome and it naturally provides isotropic accelerated expansion \cite{Horndeski:1974wa,Nicolis:2008in,Deffayet:2009wt,Deffayet:2009mn,Deffayet:2011gz}. The inclusion of non-trivial self-interactions for the scalar field has important implications in the cosmological applications \cite{DeFelice:2010pv,Kobayashi:2010cm,Kobayashi:2011nu,Gao:2011qe,Gao:2011mz,Gao:2011vs,DeFelice:2011uc,deRham:2011by,Heisenberg:2014kea}. Modifications in form of vector fields is less explored because of the difficulty of generating large scale anisotropic expansion. Nevertheless, the consideration of non-trivial self-interactions for the vector field might yield interesting cosmological consequences \cite{Horndeski:1976gi,EspositoFarese:2009aj,Jimenez:2009py,BeltranJimenez:2013fca,Jimenez:2013qsa,Jimenez:2014rna,Heisenberg:2014rta,Tasinato:2014eka,Allys:2015sht,DeFelice:2016cri,Jimenez:2016isa}.

Another interesting research line within modified gravity theories is massive gravity. Since the carrier of the gravitational force would be massive in this case, the mediated force would be Yukawa suppressed on large scales and this could yield a natural explanation of the recent cosmological acceleration. The unique mass term at the linear level was already constructed in the 1940's by Fierz and Pauli \cite{Fierz:1939ix}. This constitutes the unique linear theory at the classical level without introducing any additional ghost degree of freedom. Even theoretically being viable, this simple linear theory gives rise to the vDVZ discontinuity \cite{vanDam:1970vg,Zakharov:1970cc}, which reflects the fact that the massless limit of the theory yields a discrete difference to General Relativity. However, this discontinuity can be avoided by restoring non-linear interactions that become appreciable on small scales to freeze out the field fluctuations \cite{Vainshtein:1972sx}. The inclusion of these non-linear interactions has to be performed in a way that maintains the Boulware-Deser ghost absent \cite{Boulware:1973my}, which was finally accomplished in 2010 \cite{deRham:2010ik,deRham:2010kj,Hassan:2011vm,Hassan:2011hr,Hassan:2011zd}. Besides being the unique ghost-free non-linear theory of massive gravity, it is technically natural in the sense that it is not subject to strong renormalization by quantum loops \cite{deRham:2012ew,deRham:2013qqa}. The potential interactions have to be tuned in a very specific way to ensure the absence of the Boulware-Deser ghost, however the one-loop contributions from the gravitons usually destabilize this special structure. Notwithstanding this detuning of the potential interactions is irrelevant below the Planck scale.

Still within the same line of quantum stability of the theory, new consistent matter couplings have been proposed in \cite{deRham:2014naa,Noller:2014sta,Heisenberg:2014rka}. In order to maintain the right structure of the potential interactions at the quantum level, only very restricted matter couplings are allowed. In this context, a new non-minimal matter coupling through a very specific effective composite metric built out of the two metrics in massive (bi-)gravity was introduced. Even if this coupling reintroduces the Boulware-Deser ghost\cite{deRham:2014naa, deRham:2014fha}, it can be used as a consistent effective field theory. In fact it was argued that there is only this unique effective metric that maintains the theory ghost free up to the strong coupling scale in the metric language \cite{Huang:2015yga,Heisenberg:2015iqa}, which can be extended further in the unconstrained vielbein formulation of the theory \cite{Melville:2015dba}. Within the same formulation, it was pointed out that the ghost freedom might be preserved fully non-linearly \cite{Hinterbichler:2015yaa}. However, that this was not the case was very soon proven in \cite{deRham:2015cha}. Nevertheless, in the partially constrained vielbein formulation this can be indeed achieved at the price of losing local Lorentz symmetry \cite{DeFelice:2015yha}.
Moreover, the matter coupling through a composite effective metric has also important implications in cosmology \cite{deRham:2014naa,Enander:2014xga,Gumrukcuoglu:2014xba,Solomon:2014iwa,Gao:2014xaa,Gumrukcuoglu:2015nua,Lagos:2015sya}. Similarly, for the dark matter phenomenology in form of dipolar dark matter it plays a crucial role \cite{Blanchet:2015sra,Blanchet:2015bia,Bernard:2015gwa}. The cosmological implications of extensions of massive gravity by scalars were already investigated in \cite{Hinterbichler:2013dv,Gumrukcuoglu:2013nza,Andrews:2013uca,Bamba:2013aca,Gabadadze:2014kaa,Goon:2014ywa,Motohashi:2014una,Heisenberg:2015voa}. However, in these constructions the scalar field was living on the dynamical metric. The generalization of this to the case of doubly coupled scalar field may offer richer phenomenology and avoid some of the instabilities and fine tunings considerer in the previous models \cite{Mukohyama:2014rca,DeFelice:2015yha}.
%0264-9381-30-18-184005,

The theoretical implications of this effective composite metric along the line of non-minimally coupled scalar-tensor theories were investigated in \cite{Gao:2014xaa,Heisenberg:2015wja}. The specific model considered in \cite{Heisenberg:2015wja} is motivated by the proxy theory to massive gravity \cite{deRham:2011by}, which is constructed by covariantizing the decoupling limit of massive gravity. The non-minimal derivative couplings are this time between the scalar field and terms that depend on the Riemann tensor and Ricci tensor of the effective composite metric.
In \cite{Heisenberg:2015wja} it was investigated whether or not these non-minimal derivative couplings maintain the ghost absent in the decoupling limit and on maximally symmetric backgrounds. A detailed Hamiltonian analysis of maximally symmetric space-times showed that the only consistent non-minimal derivative coupling is through a dynamical fiducial metric but without the inclusion of an additional kinetic term. If the fiducial metric is fixed, then the Hamiltonian of maximally symmetric space-times becomes highly non-linear in the lapse, which is consistent with the perturbation analysis on top of FLRW background in \cite{Gao:2014xaa}. This is also the case for dynamical fiducial metric in the presence of an additional kinetic term. 
In this work we shall study the cosmological perturbations of these non-minimal derivative couplings in the scenario where both metrics are dynamical but without including the kinetic term of the fiducial metric.

%%%%%%%%%%%%%%%%%%%%%
\section{Dynamical composite metric} \label{sec:dyncom}

Massive gravity theories usually involve two metrics $g_{\mu\nu}$ and $f_{\mu\nu}$. A consistent coupling of some extra scalar field $\phi$ to both metrics simultaneously was introduced in \cite{deRham:2014naa} through a composite metric $\tilde{g}_{\mu\nu}$
	\begin{equation}
		\tilde{g}_{\mu\nu}\equiv\alpha^{2}g_{\mu\nu}+2\alpha\beta\,g_{\mu\lambda}X_{\phantom{\lambda}\nu}^{\lambda}+\beta^{2}f_{\mu\nu},
	\end{equation}
with $X_{\phantom{\mu}\nu}^{\mu}$ defined by
	\begin{equation}
		X_{\phantom{\mu}\lambda}^{\mu}X_{\phantom{\lambda}\nu}^{\lambda}\equiv g^{\mu\lambda}f_{\lambda\nu}.
	\end{equation}
Already in the original consideration of this coupling with the effective composite metric without derivatives acting on it, a perturbed ADM analysis divulged the existence of the Boulware-Deser ghost originated from an operator that involved spatial derivatives \cite{deRham:2014naa, deRham:2014fha}. However, the analysis performed in the decoupling limit revealed the ghost absence at the strong coupling scale $\Lambda_3$. Hence, the original massive gravity theory with this non-minimal matter coupling can be considered as an effective field theory at the very least till the strong coupling scale and subsequently its phenomenological aspects can be deduced up to this scale. In addition, the absence of the Boulware-Deser ghost was proven in the mini-superspace approximation and also around exact FLRW solution since the ghostly operators are absent in the FLRW case.
The simplest coupling between the scalar field and the composite metric is through the  kinetic term for the scalar field $\tilde{g}^{\mu\nu}\partial_{\mu}\phi\partial_{\nu}\phi$. 
It is natural to consider more general couplings which involve derivatives of the composite metric, for instance, galileon/Horndeski-like couplings. 
This possibility was investigated in \cite{Gao:2014xaa}, in which it was found that the BD ghost would arise when the metric $f_{\mu\nu}$ is considered to be fixed.
In this paper, we relax this restriction and consider bimetric theory with both dynamical metrics $g_{\mu\nu}$ and $f_{\mu\nu}$ \cite{Heisenberg:2015wja}. We consider the following action
	\begin{equation}
	S=S^{\mathrm{EH}}+S^{\mathrm{pot}}+S^{\mathrm{kin}}+S^{\mathrm{der}},\label{action}
	\end{equation}
where $S^{\mathrm{EH}}$ is the Einstein-Hilbert term for $g_{\mu\nu}$
		\begin{equation}
		S^{\mathrm{EH}}  =  \frac{M_{\mathrm{pl}}^{2}}{2}\int\!\mathrm{d}^{4}x\sqrt{-g}R\left[g\right]\,.\label{S_EH}
		\end{equation}
Note that the Einstein-Hilbert term only contains a Ricci scalar for the $g$ metric. For the viability of the model we are not allowed to consider an additional kinetic term for the $f$ metric. The non-derivative potential interactions $S^{\mathrm{pot}}$ of the two metrics are given by
		\begin{equation}
		S^{\mathrm{pot}}  =  M_{\mathrm{pl}}^{2}\int\!\mathrm{d}t\mathrm{d}^{3}x\sqrt{-g}\,m^{2}\sum_{n=0}^{4}c_{n}\,e_{n}\left(\bm{X}\right),\label{S_pot}
		\end{equation}
with $\bm{X}$ standing for $X^{\mu}_{\phantom{\mu}\nu}$ and for a matrix $M_{\phantom{\mu}\nu}^{\mu}$, $e_{n}\left(\bm{M}\right)$ are the elementary symmetric polynomials defined by
			\begin{equation}
				e_{n}\left(\bm{M}\right)\equiv n!M_{[\mu_{1}}^{\mu_{1}}M_{\mu_{2}}^{\mu_{2}}\cdots M_{\mu_{n}]}^{\mu_{n}},
			\end{equation}
where the antisymmetrization is unnormalized. We further shall consider a generalized kinetic term  $S^{\mathrm{kin}}$
for the scalar field $\phi$ with the composite metric
		\begin{equation}
		S^{\mathrm{kin}}  =  \int\!\mathrm{d}^{4}x\sqrt{-\tilde{g}}\,P\big(\tilde{X},\phi\big),\label{S_kin}
		\end{equation}
with $\tilde{X}$ denoting the kinetic term of $\phi$ in terms of the composite metric
	\begin{equation}
	\tilde{X}\equiv-\frac{1}{2}\tilde{g}^{\mu\nu}\partial_{\mu}\phi\partial_{\nu}\phi.\label{X_tilde}
	\end{equation}
The most important ingredient of our action is the non-minimal derivative coupling of the scalar field with the composite metric in $S^{\mathrm{der}}$. Motivated by the covariantization of the decoupling limit of massive gravity we shall consider the proxy theory living on the composite metric \cite{Heisenberg:2015wja},
		\begin{equation}
		S^{\mathrm{der}}  =  \int\!\mathrm{d}^{4}x\sqrt{-\tilde{g}}\left(\lambda_{1}(\phi)R\left[\tilde{g}\right]+\lambda_{2}(\phi)G^{\mu\nu}\left[\tilde{g}\right]\partial_{\mu}\phi\partial_{\nu}\phi+\lambda_{3}(\phi)L^{\mu\rho\nu\sigma}\left[\tilde{g}\right]\partial_{\mu}\phi\partial_{\nu}\phi\,\tilde{\nabla}_{\rho}\partial_{\sigma}\phi\right),\label{S_der}
		\end{equation}
which can be also thought of as the derivative interactions for the two metrics, where $R\left[\tilde{g}\right]$, $G^{\mu\nu}\left[\tilde{g}\right]$ and
		$L^{\mu\rho\nu\sigma}\left[\tilde{g}\right]$ are the Ricci scaslar,
		Einstein tensor and the double dual Riemann tensor associated with
		$\tilde{g}_{\mu\nu}$, respectively. $\tilde{\nabla}_{\mu}$ is the covariant derivative adapted to $\tilde{g}_{\mu\nu}$. The ``double dual Riemann
		tensor'' for a metric $g_{\mu\nu}$ is defined as
			\begin{equation}
			L^{\mu\rho\nu\sigma}[g]\equiv R^{\mu\rho\nu\sigma}+\left(R^{\mu\sigma}g^{\nu\rho}+R^{\nu\rho}g^{\mu\sigma}-R^{\mu\nu}g^{\rho\sigma}-R^{\rho\sigma}g^{\mu\nu}\right)+\frac{1}{2}R\left(g^{\mu\nu}g^{\rho\sigma}-g^{\mu\sigma}g^{\nu\rho}\right).\label{ddrt_def}
			\end{equation}
In the following we will study this action on FLRW background and establish our parametrization for linear perturbations.

%%%%%%%%%%%%%%%%%%%%%
\section{Cosmological parametrization}
In this section we shall determine the full set of equations of motion that dictates the evolution of spatially flat
FLRW background in the presence of an additional scalar field that couples via derivative couplings to the composite effective metric.  
After introducing our cosmological parametrization in this section, we shall analyze the linear tensor, vector and scalar perturbations on top of this background
in the next section. We shall particularly show, that the nonminimal coupling between the Horndeski field and massive gravity through the composite metric 
does not reintroduce the BD ghost.
We parametrize the two metrics $g_{\mu\nu}$ and $f_{\mu\nu}$ to be
	\begin{eqnarray}
	g_{\mu\nu}\mathrm{d}x^{\mu}\mathrm{d}x^{\nu} & = & -N^{2}\left(e^{2A}-\left(e^{-\bm{H}}\right)^{ij}B_{i}B_{j}\right)\mathrm{d}t^{2}+2NaB_{i}\mathrm{d}t\mathrm{d}x^{i}+a^{2}\left(e^{\bm{H}}\right)_{ij}\mathrm{d}x^{i}\mathrm{d}x^{j},\label{metric_g}\\
	f_{\mu\nu}\mathrm{d}x^{\mu}\mathrm{d}x^{\nu} & = & -N_{f}^{2}\left(e^{2\varphi}-\left(e^{-\bm{\Gamma}}\right)^{ij}\Omega_{i}\Omega_{j}\right)\mathrm{d}t^{2}+2N_{f}a_{f}\Omega_{i}\mathrm{d}t\mathrm{d}x^{i}+a_{f}^{2}\left(e^{\bm{\Gamma}}\right)_{ij}\mathrm{d}x^{i}\mathrm{d}x^{j},\label{metric_f}
	\end{eqnarray}
where $N$, $a$, $N_f$ and $a_f$ are functions of time only, and the matrix exponentials are defined perturbatively as $\left(e^{\bm{H}}\right)_{ij}\equiv\delta_{ij}+H_{ij}+\frac{1}{2}H_{i}^{\phantom{i}k}H_{kj}+\mathcal{O}\left(H^{3}\right)$ and $\left(e^{-\bm{H}}\right)^{ij}=\delta^{ij}-H^{ij}+\frac{1}{2}H_{\phantom{i}k}^{i}H^{kj}+\mathcal{O}\left(H^{3}\right)$, etc.
Throughout this paper, spatial indices are raised and lowered by $\delta_{ij}$ and $\delta^{ij}$.
We further decompose (with $\partial^{2}\equiv\delta^{ij}\partial_{i}\partial_{j}$)
\begin{eqnarray}
B_{i} & \equiv & \partial_{i}B+S_{i},\label{B_dec}\\
H_{ij} & \equiv & 2\zeta\,\delta_{ij}+\left(\partial_{i}\partial_{j}-\frac{1}{3}\delta_{ij}\partial^{2}\right)E+\partial_{(i}F_{j)}+h_{ij},\label{H_dec}\\
\Omega_{i} & \equiv & \partial_{i}\omega+\sigma_{i},\label{Omega_dec}\\
\Gamma_{ij} & \equiv & 2\psi\,\delta_{ij}+\left(\partial_{i}\partial_{j}-\frac{1}{3}\delta_{ij}\partial^{2}\right)\chi+\partial_{(i}\xi_{j)}+\gamma_{ij},\label{Gamma_dec}
\end{eqnarray}
with $\partial_{(i}F_{j)}\equiv\frac{1}{2}\left(\partial_{i}F_{j}+\partial_{j}F_{i}\right)$,
etc, and
\begin{equation}
\partial^{i}S_{i}=\partial^{i}F_{i}=\partial^{i}\sigma_{i}=\partial^{i}\xi_{i}=0,\qquad h_{\phantom{i}i}^{i}=\gamma_{\phantom{i}i}^{i}=0,\qquad\partial^{i}h_{ij}=\partial^{i}\gamma_{ij}=0.
\end{equation}
Accordingly, it is convenient to parametrize the composite metric to be
	\begin{equation}
		\tilde{g}_{\mu\nu}\mathrm{d}x^{\mu}\mathrm{d}x^{\nu}=-\tilde{N}^{2}\left(e^{2\tilde{A}}-(e^{-\tilde{\bm{H}}})^{ij}\tilde{B}_{i}\tilde{B}_{j}\right)\mathrm{d}t^{2}+2\tilde{N}\tilde{a}\tilde{B}_{i}\mathrm{d}t\mathrm{d}x^{i}+a^{2}(e^{\tilde{\bm{H}}})_{ij}\mathrm{d}x^{i}\mathrm{d}x^{j},
	\end{equation}
where
	\begin{equation}
		\tilde{N}\equiv\alpha\,N+\beta\,N_{f},\qquad\tilde{a}\equiv\alpha\,a+\beta\,a_{f}.
	\end{equation}
Similar to (\ref{B_dec})-(\ref{Gamma_dec}), we may also decompose
	\begin{equation}
		\tilde{B}_{i}\equiv\partial_{i}\tilde{B}+\tilde{S}_{i},\qquad\tilde{H}_{ij}\equiv2\tilde{\zeta}\,\delta_{ij}+\left(\partial_{i}\partial_{j}-\frac{1}{3}\delta_{ij}\partial^{2}\right)\tilde{E}+\partial_{(i}\tilde{F}_{j)}+\tilde{h}_{ij},
	\end{equation}
with $\partial^i \tilde{S}_i = \partial^i \tilde{F}_i = \partial^i \tilde{h}_{ij}= \delta^{ij}\tilde{h}_{ij} =0$.
Note $\tilde{A}$ etc. are expressed in terms of $\{A,B_{i},H_{ij},\varphi,\Omega_{i},\Gamma_{ij}\}$ as
	\begin{equation}
		\tilde{A}=\sum_{n=1}\tilde{A}^{(n)}\left(A,B_{i},H_{ij},\varphi,\Omega_{i},\Gamma_{ij}\right)
	\end{equation}
etc., where $n$ denotes the order in $\{A,B_{i},H_{ij},\varphi,\Omega_{i},\Gamma_{ij}\}$.
At the linear order, we have, for the scalar modes,
	\begin{eqnarray}
	\tilde{A}^{(1)} & = & \alpha\frac{N}{\tilde{N}}A+\beta\frac{N_{f}}{\tilde{N}}\varphi,\label{At1}\\
	\tilde{B}^{(1)} & = & \alpha\,r_{1}B+\beta\,r_{2}\omega,\label{Bst1}\\
	\tilde{\zeta}^{(1)} & = & \alpha\frac{a}{\tilde{a}}\zeta+\beta\frac{a_{f}}{\tilde{a}}\psi,\label{zetat1}\\
	\tilde{E}^{(1)} & = & \alpha\frac{a}{\tilde{a}}E+\beta\frac{a_{f}}{\tilde{a}}\chi,\label{Et1}
	\end{eqnarray}
with
	\begin{equation}
	r_{1}\equiv\frac{aN\left(N_{f}\tilde{a}+a_{f}\tilde{N}\right)}{\left(Na_{f}+aN_{f}\right)\tilde{a}\tilde{N}},\qquad r_{2}\equiv\frac{a_{f}N_{f}\left(N\tilde{a}+a\tilde{N}\right)}{\left(Na_{f}+aN_{f}\right)\tilde{a}\tilde{N}},\label{r1r2}
	\end{equation}
	for the vector modes,
	\begin{equation}
	\tilde{S}_{i}^{(1)}=\alpha\,r_{1}S_{i}+\beta\,r_{2}\sigma_{i},\qquad\tilde{F}_{i}^{(1)}=\alpha\frac{a}{\tilde{a}}F_{i}+\beta\frac{a_{f}}{\tilde{a}}\xi_{i},\label{SiFit1}
	\end{equation}
	and for the tensor modes
	\begin{equation}
	\tilde{h}_{ij}^{(1)}=\alpha\frac{a}{\tilde{a}}h_{ij}+\beta\frac{a_{f}}{\tilde{a}}\gamma_{ij}.\label{hijt1}
	\end{equation}
For later convenience, for any quantity $q$, we denote 
	\begin{equation}
	\dot{q} = \frac{1}{N} \frac{\mathrm{d}q}{\mathrm{d}t},\qquad q' \equiv \frac{1}{\tilde{N}} \frac{\mathrm{d}q}{\mathrm{d}t}
	\end{equation}
for short.

The background equations of motion can be determined by requiring the vanishing of the first order action of $A$, $\zeta$, $\varphi$, $\psi$ and $\delta\phi$, which is given by
	\begin{equation}
		S_{1}=\int\!\mathrm{d}t\mathrm{d}^{3}x\,Na^{3}\left(\mathcal{E}_{A}A+\mathcal{E}_{\zeta}\,3\zeta+\mathcal{E}_{\varphi}\varphi+\mathcal{E}_{\psi}3\psi+\frac{\tilde{N}\tilde{a}^{3}}{Na^{3}}\mathcal{E}_{\delta\phi}\,\delta\phi\right).
	\end{equation}
The set of equations of motion are thus given by
	\begin{eqnarray}
	\mathcal{E}_{A} & \equiv & 3M_{\mathrm{pl}}^{2}H^{2}+\mathcal{E}_{A}^{\mathrm{pot}}+\alpha\frac{\tilde{a}^{3}}{a^{3}}\mathcal{E}_{\tilde{A}}^{\mathrm{com}}=0,\label{bgeom_A}\\
	\mathcal{E}_{\zeta} & \equiv & M_{\mathrm{pl}}^{2}\left(3H^{2}+2\dot{H}\right)+\mathcal{E}_{\zeta}^{\mathrm{pot}}+\alpha\frac{\tilde{N}\tilde{a}^{2}}{Na^{2}}\mathcal{E}_{\tilde{\zeta}}^{\mathrm{com}}=0,\label{bgeom_zeta}\\
	\mathcal{E}_{\varphi} & \equiv & \mathcal{E}_{\varphi}^{\mathrm{pot}}+\beta\frac{N_{f}}{N}\frac{\tilde{a}^{3}}{a^{3}}\mathcal{E}_{\tilde{A}}^{\mathrm{com}}=0,\label{bgeom_vphi}\\
	\mathcal{E}_{\psi} & \equiv & \mathcal{E}_{\psi}^{\mathrm{pot}}+\beta\frac{a_{f}}{a}\frac{\tilde{N}\tilde{a}^{2}}{Na^{2}}\mathcal{E}_{\tilde{\zeta}}^{\mathrm{com}}=0,\label{bgeom_psi}
	\end{eqnarray}
where the Hubble parameter $H$ is defined to be $H\equiv \dot{a}/a \equiv \frac{1}{N a}\frac{\mathrm{d}a}{\mathrm{d}t}$. 
In the above,
	\begin{eqnarray}
	\mathcal{E}_{A}^{\mathrm{pot}} & = & M_{\mathrm{pl}}^{2}m^{2}\left(c_{0}+3\frac{a_{f}}{a}c_{1}+6\frac{a_{f}^{2}}{a^{2}}c_{2}+6\frac{a_{f}^{3}}{a^{3}}c_{3}\right),\label{Epot_A}\\
	\mathcal{E}_{\zeta}^{\mathrm{pot}} & = & b_{1}+\frac{aN_{f}}{Na_{f}}b_{2},\label{Epot_zeta}\\
	\mathcal{E}_{\varphi}^{\mathrm{pot}} & = & M_{\mathrm{pl}}^{2}m^{2}\frac{N_{f}}{N}\left(c_{1}+6\frac{a_{f}}{a}c_{2}+18\frac{a_{f}^{2}}{a^{2}}c_{3}+24\frac{a_{f}^{3}}{a^{3}}c_{4}\right),\label{Epot_vphi}\\
	\mathcal{E}_{\psi}^{\mathrm{pot}} & = & b_{2}+b_{3}.\label{Epot_psi}
	\end{eqnarray}
	where we have introduced
	\begin{eqnarray}
	b_{1} & \equiv & M_{\mathrm{pl}}^{2}m^{2}\left(c_{0}+2\frac{a_{f}}{a}c_{1}+2\frac{a_{f}^{2}}{a^{2}}c_{2}\right),\label{b1}\\
	b_{2} & \equiv & M_{\mathrm{pl}}^{2}m^{2}\frac{a_{f}}{a}\left(c_{1}+4\frac{a_{f}}{a}c_{2}+6\frac{a_{f}^{2}}{a^{2}}c_{3}\right),\label{b2}\\
	b_{3} & \equiv & 2M_{\mathrm{pl}}^{2}m^{2}\frac{N_{f}a_{f}}{Na}\left(c_{2}+6\frac{a_{f}}{a}c_{3}+12\frac{a_{f}^{2}}{a^{2}}c_{4}\right),\label{b3}
	\end{eqnarray}
for later convenience, and
	\begin{equation}
	\mathcal{E}_{\tilde{A}}^{\mathrm{com}}=-30\tilde{H}^{3}\tilde{X}\bar{\phi}'\lambda_{3}-18\tilde{H}^{2}\tilde{X}\lambda_{2}+6\tilde{H}^{2}\lambda_{1}+6\tilde{H}\bar{\phi}'\lambda_{1,\phi}-2\tilde{X}P_{,\tilde{X}}+P,\label{Ecom_At}
	\end{equation}
	and
	\begin{eqnarray}
	\mathcal{E}_{\tilde{\zeta}}^{\mathrm{com}} & = & P+\lambda_{1}\left(6\tilde{H}^{2}+4\tilde{H}'\right)+2\lambda_{1,\phi}(2\tilde{H}\bar{\phi}'+\bar{\phi}'')+4\tilde{X}\lambda_{1,\phi\phi}\nonumber \\
	&  & -2\lambda_{2}\left(3\tilde{H}^{2}\tilde{X}+2\tilde{H}\bar{\phi}'\bar{\phi}''+2\tilde{H}'\tilde{X}\right)-4\tilde{H}\tilde{X}\bar{\phi}'\lambda_{2,\phi}\nonumber \\
	&  & -6\tilde{H}\tilde{X}\lambda_{3}\left(2\bar{\phi}'\left(\tilde{H}^{2}+\tilde{H}'\right)+3\tilde{H}\bar{\phi}''\right)-12\tilde{H}^{2}\tilde{X}^{2}\lambda_{3,\phi}.\label{Ecom_zetat}
	\end{eqnarray}
The equation of motion for the scalar field is given by
	\begin{equation}
	\mathcal{E}_{\delta\phi}^{\mathrm{com}}=\bar{\mathcal{L}}_{,\phi}-\frac{1}{\tilde{N}\tilde{a}^{3}}\frac{\mathrm{d}}{\mathrm{d}t}\left(\tilde{a}^{3}\mathcal{J}\right),\label{Ecom_phi}
	\end{equation}
	with
	\begin{eqnarray}
	\bar{\mathcal{L}}_{,\phi} & = & P_{,\phi}+6\lambda_{1,\phi}\left(2\tilde{H}^{2}+\tilde{H}'\right)\nonumber \\
	&  & +6\lambda_{2,\phi}\left(4\tilde{H}^{2}\tilde{X}+\tilde{H}\bar{\phi}'\bar{\phi}''+\tilde{H}'\tilde{X}\right)+3\tilde{X}\lambda_{2,\phi\phi}(3\tilde{H}\bar{\phi}'+\bar{\phi}'')+2\tilde{X}^{2}\lambda_{2,\phi\phi\phi}\nonumber \\
	&  & +3\tilde{H}\tilde{X}\lambda_{3,\phi}\left(11\tilde{H}^{2}\bar{\phi}'+7\tilde{H}\bar{\phi}''+6\tilde{H}'\bar{\phi}'\right)+6\tilde{X}\lambda_{3,\phi\phi}\left(6\tilde{H}^{2}\tilde{X}+2\tilde{H}\bar{\phi}'\bar{\phi}''+\tilde{H}'\tilde{X}\right)\nonumber \\
	&  & +\frac{5}{2}\tilde{X}^{2}\lambda_{3,\phi\phi\phi}(3\tilde{H}\bar{\phi}'+\bar{\phi}'')+\tilde{X}^{3}\lambda_{3,\phi\phi\phi\phi},\label{Ecom_phi_L}
	\end{eqnarray}
	and
	\begin{eqnarray}
	\mathcal{J} & = & \bar{\phi}'P_{,\tilde{X}}+6\tilde{H}^{2}\bar{\phi}'\lambda_{2}+6\tilde{H}\tilde{X}\lambda_{2,\phi}+\tilde{X}\bar{\phi}'\lambda_{2,\phi\phi}\nonumber \\
	&  & +18\tilde{H}^{3}\tilde{X}\lambda_{3}+9\tilde{H}^{2}\tilde{X}\bar{\phi}'\lambda_{3,\phi}+6\tilde{H}\tilde{X}^{2}\lambda_{3,\phi\phi}+\frac{1}{2}\tilde{X}^{2}\bar{\phi}'\lambda_{3,\phi\phi\phi}.\label{Ecom_phi_J}
	\end{eqnarray}

%%%%%%%%%%%%%%%%%%%%%
\section{Cosmological perturbations}
Using the irreducible representation of the perturbation on top of FLRW background introduced in the previous section, we are now at the position to study the stability analysis of cosmological perturbations. We shall investigate the tensor, vector and scalar perturbations separately. We shall start with the transverse traceless part of the metric fluctuations. The quadratic action for the two tensor perturbations $h_{ij}$ and $\gamma_{ij}$ is given by
	\begin{eqnarray}
	S_{2}^{\mathrm{tensor}} & = & \frac{1}{8}\int\!\mathrm{d}t\frac{\mathrm{d}^{3}k}{\left(2\pi\right)^{3}}\bigg[Na^{3}M_{\mathrm{pl}}^{2}\left(\dot{h}_{ij}^{2}-\frac{k^{2}}{a^{2}}h_{ij}^{2}\right)+\tilde{N}\tilde{a}^{3}\left(g_{\tilde{h}\tilde{h}}\tilde{h}_{ij}'^{2}-w_{\tilde{h}\tilde{h}}\frac{k^{2}}{\tilde{a}^{2}}\tilde{h}_{ij}^{2}\right)\nonumber \\
	&  & \hspace{5em}+Na^{3}\mathcal{M}^{2}\left(h_{ij}-\gamma_{ij}\right)\left(h^{ij}-\gamma^{ij}\right)\bigg],\label{S2_ten}
	\end{eqnarray}
where $\tilde{h}_{ij}$ is a shorthand for $\tilde{h}^{(1)}_{ij}$ in (\ref{hijt1}), i.e., the linear combination of $h_{ij}$ and $\gamma_{ij}$
	\begin{equation}
		\tilde{h}_{ij} \equiv \alpha\frac{a}{\tilde{a}}h_{ij}+\beta\frac{a_{f}}{\tilde{a}}\gamma_{ij},
	\end{equation}
and furthermore we have introduced the following short-cut notations for convenience, 
	\begin{eqnarray}
	g_{\tilde{h}\tilde{h}} & = & 2\lambda_{1}-2\lambda_{2}\tilde{X}-6\lambda_{3}\tilde{H}\tilde{X}\bar{\phi}',\label{ghtht}\\
	w_{\tilde{h}\tilde{h}} & = & 2\lambda_{1}+2\lambda_{2}\tilde{X}-6\lambda_{3}\tilde{X}\bar{\phi}''-4\lambda_{3,\phi}\tilde{X}^{2},\label{whtht}
	\end{eqnarray}
together with
	\begin{equation}
	\mathcal{M}^{2}\equiv\frac{a_{f}}{a}\left[M_{\mathrm{pl}}^{2}m^{2}\left(c_{1}+2\frac{a_{f}}{a}c_{2}+2\frac{N_{f}}{N}\left(c_{2}+3\frac{a_{f}}{a}c_{3}\right)\right)+\alpha\beta\frac{\tilde{N}\tilde{a}}{Na}\mathcal{E}_{\tilde{\zeta}}^{\mathrm{com}}\right].\label{Mcal}
	\end{equation}
In difference to the standard equation for gravitational waves in General Relativity, we have an additional massive tensor mode.
The propagation speed of the tensor modes can be different from the speed of light, even though observations are quite restrictive. 
One has to impose the absence of ghost and gradient instabilities. In our case this will be equivalent to imposing $g_{\tilde{h}\tilde{h}} >0$ and $w_{\tilde{h}\tilde{h}} >0$.

Similarly, we can study the stability conditions for the vector perturbations after integrating out the non-dynamical ones. The quadratic action for the four vector modes $S_i$, $F_i$, $\sigma_i$ and $\xi_i$ is given by
	\begin{align}
	S_{2}^{\mathrm{vector}}  = & \int\!\mathrm{d}t\frac{\mathrm{d}^{3}k}{\left(2\pi\right)^{3}}\bigg[\frac{1}{4}Na^{3}\,M_{\mathrm{pl}}^{2}k^{2}\left(\frac{1}{a}S_{i}-\frac{1}{2}\dot{F}_{i}\right)^{2}+\frac{1}{4}\tilde{N}\tilde{a}^{3}\,g_{\tilde{h}\tilde{h}}k^{2}\left(\frac{1}{\tilde{a}}\tilde{S}_{i}-\frac{1}{2}\tilde{F}_{i}'\right)^{2}\nonumber \\
	  & \hspace{4em}-\frac{1}{2}Na^{3}\mathcal{C}\left(S_{i}-\frac{aN_{f}}{Na_{f}}\sigma_{i}\right)^2 +\frac{Na^{3}}{16}\mathcal{M}^{2}k^{2}\left(F_{i}-\xi_{i}\right)^2\bigg],\label{S2_vec}
	\end{align}
where similar to $\tilde{h}_{ij}$ in (\ref{S2_ten}), $\tilde{S}_{i}$ and $\tilde{F}_{i}$ should be understood as the linear combinations of $S_i$ and $\sigma_i$, $F_i$ and $\xi_i$ through (\ref{SiFit1}), respectively. 
In (\ref{S2_vec}), $\mathcal{M}^2$ is given in (\ref{Mcal}) and we also introduce
	\begin{equation}
	\mathcal{C}\equiv\frac{1}{1+\frac{aN_{f}}{Na_{f}}}b_{2}+\frac{\alpha\beta}{\left(1+\frac{aN_{f}}{Na_{f}}\right)^{2}}\frac{\tilde{N}\tilde{a}a_{f}}{Na^{2}}\left(\Big(1+\frac{\tilde{a}N_{f}}{\tilde{N}a_{f}}+\frac{N\tilde{a}}{\tilde{N}a}\Big)\mathcal{E}_{\tilde{A}}^{\mathrm{com}}-\mathcal{E}_{\tilde{\zeta}}^{\mathrm{com}}\right) \label{Ccal}
	\end{equation}
with $b_2$ given in (\ref{b2}) for short. Since the vector modes $S_i$ and $\sigma_i$ have no dynamics in (\ref{S2_vec}), we may solve them in terms of $F_i$ and $\xi_i$ and arrive at the reduced action for $F_i$ and $\xi_i$, which is given by
	\begin{equation}
		S_{2}^{\mathrm{vector}}=\frac{1}{16}\int\!\mathrm{d}t\frac{\mathrm{d}^{3}k}{\left(2\pi\right)^{3}}Na^{3}k^{2}\left\{ \mathcal{G}_{\mathrm{v}}\left[\frac{1}{N}\frac{\mathrm{d}}{\mathrm{d}t}\left(\frac{a_{f}}{\tilde{a}}\left(F_{i}-\xi_{i}\right)\right)\right]^{2}+\mathcal{M}^{2}\left(F_{i}-\xi_{i}\right)^{2}\right\} , \label{S2_vec_f}
	\end{equation}
where 
	\begin{equation}
		\mathcal{G}_{\mathrm{v}}\equiv\frac{\beta^{2}M_{\text{pl}}^{2}}{1-\frac{\tilde{N}}{N}M_{\text{pl}}^{2}\left(\beta^{2}\frac{N\tilde{N}a_{f}^{2}r_{2}^{2}}{2\mathcal{C}N_{f}^{2}\tilde{a}^{2}a^{2}}k^{2}-\frac{a^{3}}{g_{\tilde{h}\tilde{h}}\tilde{a}^{3}}\right)}.
	\end{equation}
From (\ref{S2_vec_f}) it is transparent that there are two vectorial degrees of freedom giving that $\beta\neq 0$, which can be identified as $F_i - \xi_i$. For the stability condition we have to impose $\mathcal{G}_{\mathrm{v}}>0$.

Last but not least we study now the linear stability of the scalar modes in our model.
Initially we have 9 scalar modes, of which four ($A$, $B$, $\zeta$ and $E$) are from $g_{\mu\nu}$, four ($\varphi$, $\omega$, $\psi$ and $\chi$) are from $f_{\mu\nu}$, and one is the perturbation of the scalar field $\delta\phi$.
In order to simplify the calculation, we choose a gauge in which $\delta\phi = E =0$.
In the residual 7 modes, only 2 modes are dynamical, which can be conveniently chosen to be $\tilde{\zeta}^{(1)}$ as in (\ref{zetat1}) and $\tilde{E}^{(1)}$ with $E=0$ as in (\ref{Et1}), respectively.
After some manipulations, the final quadratic action for these two scalar modes takes the following general structure
	\begin{eqnarray}
	S_{2}^{\mathrm{scalar}} & = & \int\mathrm{d}t\frac{\mathrm{d}^{3}k}{\left(2\pi\right)^{3}}\bigg[\mathcal{G}_{11}(\partial_{t}\tilde{\zeta})^{2}+2\mathcal{G}_{12}\partial_{t}\tilde{\zeta}\partial_{t}\tilde{E}+\mathcal{G}_{22}(\partial_{t}\tilde{E})^{2}\nonumber \\
	&  & \hspace{5em} +\Xi\left(\partial_{t}\tilde{\zeta}\,\tilde{E}-\tilde{\zeta}\,\partial_{t}\tilde{E}\right)+\mathcal{W}_{11}\tilde{\zeta}^{2}+2\mathcal{W}_{12}\tilde{\zeta}\tilde{E}+\mathcal{W}_{22}\tilde{E}^{2}\bigg].\label{S2s_fin}
	\end{eqnarray}
The explicit expressions for the coefficients $\mathcal{G}_{11}$ etc. are too involved to be presented in the main text. In Appendix \ref{app:sca} we describe the details in deriving (\ref{S2s_fin}) as well as the full expressions for the coefficients. As one can see, the would-be Boulware-Deser ghost does not propagate and could be integrated out. Only two scalar fields are dynamical, one of which comes from the matter field itself. This is in complete agreement with the findings in \cite{Heisenberg:2015wja}. The hamiltonian analysis in the mini-superspace in \cite{Heisenberg:2015wja} revealed that non-minimal derivate couplings in massive bigravity with both metrics being dynamical and without the presence of the kinetic term for the $f$ metric does not reintroduce the Boulware-Deser ghost below the strong coupling scale. Our cosmological perturbations analysis here affirmed the same conclusion and is supplementary to the analysis in \cite{Heisenberg:2015wja}.

%%%%%%%%%%%%%%%%%%%%%
\section{Conclusion}
This paper was dedicated to the detail analysis of cosmological perturbations in massive bigravity in the presence of non-minimal derivative couplings of the composite effective metric to a Horndeski scalar field. After working out the governing background equations of motion we payed special attention to the stability analysis of tensor, vector and scalar perturbations. The tensor perturbations showed the presence of four propagating tensor modes and we have seen which dynamical quantities have to be constrained in order to avoid ghost and gradient instabilities. Similarly, the analysis of vector perturbations disclosed that out of naively counted four vector modes only two of them are actually dynamical. The absence of ghost and gradient instabilities required this time the condition $\mathcal{G}_{\mathrm{v}}>0$ on the background dynamics. Finally, the analysis of scalar perturbations resulted in two propagating scalar modes and hence reinforcing the fact that the Boulware-Deser ghost is not excited. 

In this work, we considered a uniform composite metric $\tilde{g}_{\mu\nu}$ in (\ref{S_kin}) and (\ref{S_der}) for simplicity. 
It would be interesting to investigate the case with different composite metrics for each non-minimal derivative coupling terms, such as those in \cite{Gao:2014xaa}.
A special case is just the bigravity with composite metric appearing only in the kinetic term of the scalar field $\tilde{X}$ \cite{Gumrukcuoglu:2015nua}, which yields a viable theory.
On the other hand, in light of the analysis in \cite{Heisenberg:2015wja}, the healthiness of theories with different composite metrics for each non-minimal coupling terms still remains an open question.

% % % % % % % % % % % % % %

\acknowledgments

X.G. was supported by JSPS Grant-in-Aid for Scientific Research No. 25287054 and 26610062.
LH acknowledges financial support from Dr. Max R\"ossler, the Walter Haefner Foundation and the ETH Zurich Foundation.

%%%%%%%%%%
\appendix

\section{Details in quadratic action for the scalar modes} \label{app:sca}

In this appendix we provide the details in deriving (\ref{S2s_fin}).
After fixing the gauge $\delta\phi= E =0$, in the quadratic action for the residual 7 scalar modes $S_{2}[A,B,\varphi,\omega,\zeta,\psi,\chi]$, four modes $A$, $B$, $\varphi$ and $\omega$ have no explicit time derivatives and thus can be integrated out.
We thus get the quadratic action for 3 variables $\{\zeta,\psi,\chi\}$ or more conveniently $\{\zeta,\tilde{\zeta},\tilde{E}\}$, where $\tilde{a}$ and $\tilde{E}$ are given in (\ref{zetat1}) and (\ref{Et1}) after setting $E=0$.
The quadratic action is given by
	\begin{eqnarray}
	S_{2}[\zeta,\tilde{\zeta},\tilde{E}] & = & \int\!\mathrm{d}t\frac{\mathrm{d}^{3}k}{\left(2\pi\right)^{3}}\,\frac{1}{\Delta}\Big[\mathcal{A}_{1}\,\zeta^{2}+\mathcal{A}_{2}\,\zeta\tilde{\zeta}+\mathcal{A}_{3}\,\zeta\tilde{E}+\mathcal{A}_{4}\,\zeta\partial_{t}\tilde{\zeta}+\mathcal{A}_{5}\,\zeta\partial_{t}\tilde{E}+\mathcal{A}_{6}\,\tilde{\zeta}^{2}\nonumber \\
	&  & \hspace{4em}+\mathcal{A}_{7}\,\tilde{\zeta}\,\tilde{E}+\mathcal{A}_{8}\,\tilde{\zeta}\partial_{t}\tilde{\zeta}+\mathcal{A}_{9}\,\tilde{\zeta}(\partial_{t}\tilde{E})+\mathcal{A}_{10}\,\tilde{E}^{2}+\mathcal{A}_{11}\,\tilde{E}\,\partial_{t}\tilde{\zeta}\nonumber \\
	&  & \hspace{4em}+\mathcal{A}_{12}\,\tilde{E}\,\partial_{t}\tilde{E}+\mathcal{A}_{13}\,(\partial_{t}\tilde{\zeta})^{2}+\mathcal{A}_{14}\,\partial_{t}\tilde{\zeta}\,\partial_{t}\tilde{E}+\mathcal{A}_{15}\,(\partial_{t}\tilde{E})^{2}\Big],\label{S2_s_3v} 
	\end{eqnarray}
where we defined
	\begin{equation}
	\Delta\equiv C_{1}\left(M_{\text{pl}}^{2}(2m_{\tilde{A}\tilde{A}}+\mathcal{E}_{\tilde{A}}^{\text{com}})+6\frac{N\tilde{a}^{3}}{\tilde{N}a^{3}}\tilde{H}^{2}f_{\tilde{A}\tilde{\zeta}}^{2}\right)+4M_{\text{pl}}^{2}\frac{k^{2}}{a^{2}}\beta^{2}\frac{N\tilde{a}^{3}}{\tilde{N}a^{3}}\tilde{H}^{2}f_{\tilde{A}\tilde{\zeta}}^{2}.\label{Del_def}
	\end{equation}
and coefficients $\mathcal{A}_{1},\cdots,\mathcal{A}_{15}$ in (\ref{S2_s_3v}) are given by
{\small
	\begin{eqnarray}
	\mathcal{A}_{1} & = & 2\frac{\tilde{a}^{3}N^{2}}{\tilde{N}}f_{\tilde{A}\tilde{\zeta}}^{2}M_{\text{pl}}^{2}\frac{\tilde{H}^{2}}{H^{2}}\frac{k^{4}}{a^{4}}\left(2\beta^{2}\dot{H}M_{\text{pl}}^{2}+C_{1}\right)\nonumber \\
	&  & +\frac{a^{3}N^{2}}{\tilde{N}}M_{\text{pl}}^{2}\frac{1}{\beta^{2}}\frac{k^{2}}{a^{2}}\Bigg\{6\frac{\tilde{H}}{H}C_{1}f_{\tilde{A}\tilde{\zeta}}\frac{\tilde{a}^{3}}{a^{3}}\bigg(\frac{\tilde{H}}{H}C_{2}f_{\tilde{A}\tilde{\zeta}}-M_{\text{pl}}^{2}m^{2}\frac{\tilde{N}a^{3}}{N\tilde{a}^{3}}C_{3}\bigg)+\beta^{2}\frac{\tilde{N}}{N}\Bigg[\frac{\dot{H}}{H^{2}}C_{1}M_{\text{pl}}^{2}(2m_{\tilde{A}\tilde{A}}+\mathcal{E}_{\tilde{A}}^{\text{com}})\nonumber \\
	&  & +6\frac{N\tilde{a}^{3}}{\tilde{N}a^{3}}\tilde{H}^{2}f_{\tilde{A}\tilde{\zeta}}^{2}\Bigg(C_{5}-M_{\text{pl}}^{2}\bigg(3C_{2}-\frac{\dot{H}}{H^{2}}C_{2}+\frac{\dot{C}_{2}}{H}\bigg)+\frac{\dot{H}}{H^{2}}C_{1}\Bigg)\Bigg]\Bigg\}\nonumber \\
	&  & +\frac{3}{2}\frac{a^{6}\tilde{N}}{\tilde{a}^{3}}\frac{1}{\beta^{4}}C_{1}\Bigg[3\left(m^{2}C_{3}M_{\text{pl}}^{3}-\frac{N\tilde{a}^{3}\tilde{H}}{a^{3}H\tilde{N}}C_{2}f_{\tilde{A}\tilde{\zeta}}M_{\text{pl}}\right)^{2}\nonumber \\
	&  & +\beta^{2}\frac{N\tilde{a}^{3}}{\tilde{N}a^{3}}\Bigg(C_{5}+M_{\text{pl}}^{2}\bigg(\frac{\dot{H}}{H^{2}}C_{2}-3C_{2}-\frac{\dot{C}_{2}}{H}\bigg)\Bigg)\bigg(M_{\text{pl}}^{2}(2m_{\tilde{A}\tilde{A}}+\mathcal{E}_{\tilde{A}}^{\text{com}})+6\frac{N\tilde{a}^{3}}{\tilde{N}a^{3}}\tilde{H}^{2}f_{\tilde{A}\tilde{\zeta}}^{2}\bigg)\Bigg],\label{Acal1}
	\end{eqnarray}
	\begin{eqnarray}
	\mathcal{A}_{2} & = & -3a^{3}\tilde{N}\frac{1}{\beta^{4}}C_{1}\Bigg\{-3M_{\text{pl}}^{2}\left(C_{4}-m^{2}\frac{N\tilde{a}}{\tilde{N}a}\frac{\tilde{H}}{H}C_{3}f_{\tilde{A}\tilde{\zeta}}\right)\left(m^{2}C_{3}M_{\text{pl}}^{2}-\frac{N\tilde{a}^{3}}{\tilde{N}a^{3}}\frac{\tilde{H}}{H}C_{2}f_{\tilde{A}\tilde{\zeta}}\right)\nonumber \\
	&  & +\beta^{2}\frac{N}{\tilde{N}}\bigg(M_{\text{pl}}^{2}(2m_{\tilde{A}\tilde{A}}+\mathcal{E}_{\tilde{A}}^{\text{com}})+6\frac{N\tilde{a}^{3}}{\tilde{N}a^{3}}\tilde{H}^{2}f_{\tilde{A}\tilde{\zeta}}^{2}\bigg)\bigg[C_{6}+m^{2}\frac{\tilde{a}}{a}M_{\text{pl}}^{2}\bigg(\Big(2-\frac{\dot{H}}{H^{2}}+\frac{\tilde{H}}{H}\frac{\tilde{N}}{N}\Big)C_{3}+\frac{\dot{C}_{3}}{H}\bigg)\bigg]\Bigg\}\nonumber \\
	&  & -6M_{\text{pl}}^{2}\frac{k^{2}}{a^{2}}\frac{a^{5}\tilde{N}}{\tilde{a}^{2}}\frac{1}{\beta^{2}}\Bigg\{ C_{1}\bigg[-m^{2}C_{3}\bigg(g_{\tilde{h}\tilde{h}}M_{\text{pl}}^{2}+\frac{N^{2}\tilde{a}^{6}}{\tilde{N}^{2}a^{6}}\frac{\tilde{H}^{2}}{H^{2}}f_{\tilde{A}\tilde{\zeta}}^{2}\bigg)+\frac{N\tilde{a}^{3}}{\tilde{N}a^{3}}\frac{\tilde{H}}{H}f_{\tilde{A}\tilde{\zeta}}\left(C_{2}g_{\tilde{h}\tilde{h}}+\frac{\tilde{a}^{2}}{a^{2}}C_{4}\right)\bigg]\nonumber \\
	&  & +2\beta^{2}\frac{N^{2}\tilde{a}^{5}}{\tilde{N}^{2}a^{5}}\tilde{H}^{2}f_{\tilde{A}\tilde{\zeta}}^{2}\bigg[C_{6}+M_{\text{pl}}^{2}m^{2}\frac{\tilde{a}}{a}\bigg(\Big(2-\frac{\dot{H}}{H^{2}}+\frac{\tilde{N}\tilde{H}}{NH}\Big)C_{3}+\frac{\dot{C}_{3}}{H}\bigg)\bigg]\Bigg\}\nonumber \\
	&  & -4M_{\text{pl}}^{2}\frac{k^{4}}{a^{2}}\tilde{a}N\frac{\tilde{H}}{H}C_{1}f_{\tilde{A}\tilde{\zeta}}g_{\tilde{h}\tilde{h}},
	\end{eqnarray}
	\begin{equation}
	\mathcal{A}_{3}=k^{4}M_{\text{pl}}^{2}\tilde{a}NC_{1}g_{\tilde{h}\tilde{h}}\bigg[\frac{1}{\beta^{2}}\Big(m^{2}M_{\text{pl}}^{2}\frac{\tilde{N}a^{3}}{N\tilde{a}^{3}}C_{3}-\frac{\tilde{H}}{H}C_{2}f_{\tilde{A}\tilde{\zeta}}\Big)-\frac{2}{3}\frac{k^{2}}{a^{2}}\frac{\tilde{H}}{H}f_{\tilde{A}\tilde{\zeta}}\bigg],
	\end{equation}
	\begin{eqnarray}
	\mathcal{A}_{4} & = & 2\frac{k^{2}}{a^{2}}\frac{\tilde{a}^{3}N}{\tilde{N}}\frac{M_{\text{pl}}^{2}}{H}\bigg[6M_{\text{pl}}^{2}m^{2}C_{3}H\tilde{H}f_{\tilde{A}\tilde{\zeta}}\Big(g_{\tilde{h}\tilde{h}}-\frac{\tilde{a}}{a}\frac{\tilde{H}}{H}f_{\tilde{A}\tilde{\zeta}}\Big)+C_{1}\Big(g_{\tilde{h}\tilde{h}}(2m_{\tilde{A}\tilde{A}}+\mathcal{E}_{\tilde{A}}^{\text{com}})-6\tilde{H}^{2}f_{\tilde{A}\tilde{\zeta}}^{2}\Big)\bigg]\nonumber \\
	&  & +3\frac{\tilde{a}^{3}N}{\tilde{N}}C_{1}\frac{M_{\text{pl}}^{2}}{\beta^{2}H}\bigg[m^{2}C_{3}\bigg(6H\tilde{H}f_{\tilde{A}\tilde{\zeta}}\Big(\frac{\tilde{N}a^{3}}{N\tilde{a}^{3}}M_{\text{pl}}^{2}+g_{\tilde{h}\tilde{h}}\Big)-6\frac{\tilde{a}}{a}\tilde{H}^{2}f_{\tilde{A}\tilde{\zeta}}^{2}-\frac{\tilde{N}a^{2}}{N\tilde{a}^{2}}M_{\text{pl}}^{2}(2m_{\tilde{A}\tilde{A}}+\mathcal{E}_{\tilde{A}}^{\text{com}})\bigg)\nonumber \\
	&  & \hspace{4em}+C_{2}\left(g_{\tilde{h}\tilde{h}}(2m_{\tilde{A}\tilde{A}}+\mathcal{E}_{\tilde{A}}^{\text{com}})-6\tilde{H}^{2}f_{\tilde{A}\tilde{\zeta}}^{2}\right)\bigg],
	\end{eqnarray}
	\begin{eqnarray}
	\mathcal{A}_{5} & = & \frac{N\tilde{a}^{3}}{\tilde{N}}g_{\tilde{h}\tilde{h}}\frac{M_{\text{pl}}^{2}}{2H}k^{2}\bigg[C_{1}\frac{1}{\beta^{2}}\left(6m^{2}H\tilde{H}C_{3}f_{\tilde{A}\tilde{\zeta}}+C_{2}(2m_{\tilde{A}\tilde{A}}+\mathcal{E}_{\tilde{A}}^{\text{com}})\right)\nonumber \\
	&  & \hspace{4em}+\frac{2}{3}\frac{k^{2}}{a^{2}}\left(6m^{2}H\tilde{H}C_{3}f_{\tilde{A}\tilde{\zeta}}M_{\text{pl}}^{2}+C_{1}(2m_{\tilde{A}\tilde{A}}+\mathcal{E}_{\tilde{A}}^{\text{com}})\right)\bigg],
	\end{eqnarray}
	\begin{eqnarray}
	\mathcal{A}_{6} & = & \frac{3}{2}\tilde{a}^{3}\tilde{N}\frac{1}{\beta^{2}}C_{1}\bigg[\frac{aN}{\tilde{a}\tilde{N}}C_{7}\Big(M_{\text{pl}}^{2}(2m_{\tilde{A}\tilde{A}}+\mathcal{E}_{\tilde{A}}^{\text{com}})+6\frac{N\tilde{a}^{3}}{\tilde{N}a^{3}}\tilde{H}^{2}f_{\tilde{A}\tilde{\zeta}}^{2}\Big)+3\frac{1}{\beta^{2}}M_{\text{pl}}^{2}\Big(C_{4}-m^{2}\frac{N\tilde{a}}{\tilde{N}a}\frac{\tilde{H}}{H}C_{3}f_{\tilde{A}\tilde{\zeta}}\Big)^{2}\bigg]\nonumber \\
	&  & +k^{2}\tilde{a}\tilde{N}\bigg[\Big(M_{\text{pl}}^{2}(2m_{\tilde{A}\tilde{A}}+\mathcal{E}_{\tilde{A}}^{\text{com}})+6\frac{\tilde{a}^{3}N}{a^{3}\tilde{N}}\tilde{H}^{2}f_{\tilde{A}\tilde{\zeta}}^{2}\Big)C_{1}w_{\tilde{h}\tilde{h}}+6\frac{N^{2}\tilde{a}^{4}}{\tilde{N}^{2}a^{4}}\tilde{H}^{2}C_{7}f_{\tilde{A}\tilde{\zeta}}^{2}M_{\text{pl}}^{2}\nonumber \\
	&  & +6M_{\text{pl}}^{2}\frac{1}{\beta^{2}}C_{1}g_{\tilde{h}\tilde{h}}\Big(C_{4}-m^{2}\frac{\tilde{a}N}{a\tilde{N}}\frac{\tilde{H}}{H}C_{3}f_{\tilde{A}\tilde{\zeta}}\Big)\bigg]+2k^{4}\frac{M_{\text{pl}}^{2}}{\tilde{a}}\tilde{N}\Big(C_{1}g_{\tilde{h}\tilde{h}}^{2}+2\beta^{2}\frac{N\tilde{a}^{5}}{\tilde{N}a^{5}}\tilde{H}^{2}f_{\tilde{A}\tilde{\zeta}}^{2}w_{\tilde{h}\tilde{h}}\Big),
	\end{eqnarray}
	\begin{eqnarray}
	\mathcal{A}_{7} & = & k^{4}\tilde{a}\tilde{N}C_{1}\bigg[\frac{1}{3}w_{\tilde{h}\tilde{h}}\Big(M_{\text{pl}}^{2}(2m_{\tilde{A}\tilde{A}}+\mathcal{E}_{\tilde{A}}^{\text{com}})+6\frac{N\tilde{a}^{3}}{\tilde{N}a^{3}}\tilde{H}^{2}f_{\tilde{A}\tilde{\zeta}}^{2}\Big)+\frac{1}{\beta^{2}}g_{\tilde{h}\tilde{h}}M_{\text{pl}}^{2}\Big(C_{4}-m^{2}\frac{N\tilde{a}}{\tilde{N}a}\frac{\tilde{H}}{H}C_{3}f_{\tilde{A}\tilde{\zeta}}\Big)\bigg]\nonumber \\
	&  & +\frac{2}{3}M_{\text{pl}}^{2}k^{6}\frac{\tilde{N}}{\tilde{a}}\left(C_{1}g_{\tilde{h}\tilde{h}}^{2}+2\beta^{2}\frac{N\tilde{a}^{5}}{\tilde{N}a^{5}}\tilde{H}^{2}f_{\tilde{A}\tilde{\zeta}}^{2}w_{\tilde{h}\tilde{h}}\right),
	\end{eqnarray}
	\begin{eqnarray}
	\mathcal{A}_{8} & = & 8k^{4}\beta^{2}\frac{N\tilde{a}^{4}}{\tilde{N}a^{5}}\tilde{H}f_{\tilde{A}\tilde{\zeta}}g_{\tilde{h}\tilde{h}}^{2}M_{\text{pl}}^{2}+12k^{2}\tilde{a}\tilde{H}f_{\tilde{A}\tilde{\zeta}}g_{\tilde{h}\tilde{h}}\bigg[C_{1}\Big(M_{\text{pl}}^{2}+\frac{N\tilde{a}^{3}}{\tilde{N}a^{3}}g_{\tilde{h}\tilde{h}}\Big)+\frac{N\tilde{a}^{5}}{\tilde{N}a^{5}}C_{4}M_{\text{pl}}^{2}\bigg]\nonumber \\
	&  & +3\tilde{a}^{3}\frac{1}{\beta^{2}}C_{1}\bigg[6\tilde{H}C_{4}f_{\tilde{A}\tilde{\zeta}}\Big(M_{\text{pl}}^{2}+\frac{N\tilde{a}^{3}}{\tilde{N}a^{3}}g_{\tilde{h}\tilde{h}}\Big)\nonumber \\
	&  & +\frac{N\tilde{a}}{\tilde{N}a}m^{2}\frac{1}{H}C_{3}M_{\text{pl}}^{2}\left(g_{\tilde{h}\tilde{h}}(2m_{\tilde{A}\tilde{A}}+\mathcal{E}_{\tilde{A}}^{\text{com}})-6\tilde{H}^{2}f_{\tilde{A}\tilde{\zeta}}^{2}\right)\bigg],
	\end{eqnarray}
	\begin{eqnarray}
	\mathcal{A}_{9} & = & \frac{1}{2}k^{2}\frac{N\tilde{a}^{4}}{\tilde{N}a}C_{1}g_{\tilde{h}\tilde{h}}\frac{1}{\beta^{2}}\left[m^{2}C_{3}\frac{1}{H}M_{\text{pl}}^{2}(2m_{\tilde{A}\tilde{A}}+\mathcal{E}_{\tilde{A}}^{\text{com}})+6\frac{\tilde{a}^{2}}{a^{2}}\tilde{H}C_{4}f_{\tilde{A}\tilde{\zeta}}\right]\nonumber \\
	&  & +2k^{4}\frac{N\tilde{a}^{4}}{\tilde{N}a^{3}}\tilde{H}f_{\tilde{A}\tilde{\zeta}}g_{\tilde{h}\tilde{h}}\left(C_{1}g_{\tilde{h}\tilde{h}}+\frac{\tilde{a}^{2}}{a^{2}}C_{4}M_{\text{pl}}^{2}\right)+\frac{4}{3}\beta^{2}k^{6}\frac{N\tilde{a}^{4}}{\tilde{N}a^{5}}\tilde{H}f_{\tilde{A}\tilde{\zeta}}g_{\tilde{h}\tilde{h}}^{2}M_{\text{pl}}^{2},
	\end{eqnarray}
	\begin{eqnarray}
	\mathcal{A}_{10} & = & k^{8}\tilde{N}^{2}\frac{M_{\text{pl}}^{2}}{18\tilde{a}\tilde{N}}\left(C_{1}g_{\tilde{h}\tilde{h}}^{2}+2\beta^{2}\frac{N\tilde{a}^{5}}{\tilde{N}a^{5}}\tilde{H}^{2}f_{\tilde{A}\tilde{\zeta}}^{2}w_{\tilde{h}\tilde{h}}\right)\nonumber \\
	&  & +\frac{1}{36}k^{6}\tilde{a}\tilde{N}\left[C_{1}w_{\tilde{h}\tilde{h}}\left(M_{\text{pl}}^{2}(2m_{\tilde{A}\tilde{A}}+\mathcal{E}_{\tilde{A}}^{\text{com}})+6\frac{N\tilde{a}^{3}}{\tilde{N}a^{3}}\tilde{H}^{2}f_{\tilde{A}\tilde{\zeta}}^{2}\right)+12\frac{N^{2}\tilde{a}^{4}}{\tilde{N}^{2}a^{4}}\tilde{H}^{2}C_{8}f_{\tilde{A}\tilde{\zeta}}^{2}M_{\text{pl}}^{2}\right]\nonumber \\
	&  & +\frac{1}{12}k^{4}\frac{1}{\beta^{2}}\frac{N\tilde{a}^{2}}{\tilde{N}a^{2}}C_{1}C_{8}\left(a^{3}\tilde{N}M_{\text{pl}}{}^{2}(2m_{\tilde{A}\tilde{A}}+\mathcal{E}_{\tilde{A}}^{\text{com}})+6N\tilde{a}^{3}\tilde{H}^{2}f_{\tilde{A}\tilde{\zeta}}^{2}\right),
	\end{eqnarray}
	\begin{equation}
	\mathcal{A}_{11}=2k^{4}\tilde{a}\tilde{H}f_{\tilde{A}\tilde{\zeta}}g_{\tilde{h}\tilde{h}}\left[C_{1}\left(M_{\text{pl}}^{2}+\frac{N\tilde{a}^{3}}{\tilde{N}a^{3}}g_{\tilde{h}\tilde{h}}\right)+\frac{2}{3}M_{\text{pl}}^{2}\beta^{2}\frac{k^{2}}{a^{2}}\frac{N\tilde{a}^{3}}{\tilde{N}a^{3}}g_{\tilde{h}\tilde{h}}\right],
	\end{equation}
	\begin{equation}
	\mathcal{A}_{12}=\frac{1}{3}k^{6}\frac{N\tilde{a}^{4}}{\tilde{N}a^{3}}\left(C_{1}+\frac{2}{3}\beta^{2}\frac{k^{2}}{a^{2}}M_{\text{pl}}^{2}\right)\tilde{H}g_{\tilde{h}\tilde{h}}^{2}f_{\tilde{A}\tilde{\zeta}},
	\end{equation}
	\begin{equation}
	\mathcal{A}_{13}=-3\frac{\tilde{a}^{3}}{\tilde{N}}\left(g_{\tilde{h}\tilde{h}}(2m_{\tilde{A}\tilde{A}}+\mathcal{E}_{\tilde{A}}^{\text{com}})-6\tilde{H}^{2}f_{\tilde{A}\tilde{\zeta}}^{2}\right)\left[C_{1}\left(M_{\text{pl}}^{2}+\frac{N\tilde{a}^{3}}{\tilde{N}a^{3}}g_{\tilde{h}\tilde{h}}\right)+\frac{2}{3}M_{\text{pl}}^{2}\beta^{2}\frac{k^{2}}{a^{2}}\frac{N\tilde{a}^{3}}{\tilde{N}a^{3}}g_{\tilde{h}\tilde{h}}\right],
	\end{equation}
	\begin{equation}
	\mathcal{A}_{14}=k^{2}\frac{N\tilde{a}^{6}}{\tilde{N}^{2}a^{3}}g_{\tilde{h}\tilde{h}}\left(C_{1}+\frac{2}{3}\beta^{2}\frac{k^{2}}{a^{2}}M_{\text{pl}}^{2}\right)\left(6\tilde{H}^{2}f_{\tilde{A}\tilde{\zeta}}^{2}-g_{\tilde{h}\tilde{h}}(2m_{\tilde{A}\tilde{A}}+\mathcal{E}_{\tilde{A}}^{\text{com}})\right),
	\end{equation}
	\begin{eqnarray}
	\mathcal{A}_{15} & = & \frac{1}{36}k^{4}\frac{\tilde{a}^{3}}{\tilde{N}}g_{\tilde{h}\tilde{h}}\bigg[3C_{1}M_{\text{pl}}^{2}(2m_{\tilde{A}\tilde{A}}+\mathcal{E}_{\tilde{A}}^{\text{com}})\nonumber \\
	&  & +3\frac{N\tilde{a}^{3}}{\tilde{N}a^{3}}\left(C_{1}+\frac{2}{3}\beta^{2}\frac{k^{2}}{a^{2}}M_{\text{pl}}^{2}\right)\left(6\tilde{H}^{2}f_{\tilde{A}\tilde{\zeta}}^{2}-g_{\tilde{h}\tilde{h}}(2m_{\tilde{A}\tilde{A}}+\mathcal{E}_{\tilde{A}}^{\text{com}})\right)\bigg].\label{Acal15}
	\end{eqnarray}
	}
In the above, $C_1,\cdots,C_8$ are given by
	\begin{eqnarray}
	C_{1} & = & \frac{\tilde{a}^{3}}{a_{f}^{2}a}\frac{1}{\left(1+\frac{\tilde{N}a}{N\tilde{a}}\right)^{2}}\bigg[\alpha\beta\bigg(\mathcal{E}_{\tilde{\zeta}}^{\text{com}}\frac{\tilde{N}a_{f}}{aN}-\mathcal{E}_{\tilde{A}}^{\text{com}}\Big(\frac{a_{f}\tilde{N}}{aN}+\frac{a_{f}\tilde{a}}{a^{2}}+\frac{N_{f}\tilde{a}}{aN}\Big)\bigg)\nonumber \\
	&  & \hspace{5em}-M_{\text{pl}}^{2}m^{2}\bigg(6c_{3}\frac{a_{f}^{2}}{a^{2}}+4c_{2}\frac{a_{f}}{a}+c_{1}\bigg)\left(\frac{N_{f}}{N}\frac{a}{\tilde{a}}+\frac{a_{f}}{\tilde{a}}\right)\bigg],\label{C1}
	\end{eqnarray}
	\begin{eqnarray}
	C_{2} & = & 3\beta^{2}H^{2}+m^{2}\bigg[2\left(12c_{4}\alpha^{2}+\beta(c_{2}\beta-6c_{3}\alpha)\right)\frac{a_{f}^{2}}{a^{2}}\nonumber \\
	&  & +2\left(6c_{3}\alpha^{2}+\beta(c_{1}\beta-4c_{2}\alpha)\right)\frac{a_{f}}{a}+\left(2c_{2}\alpha^{2}+\beta(c_{0}\beta-2c_{1}\alpha)\right)\bigg],
	\end{eqnarray}
	\begin{equation}
	C_{3}=6(c_{3}\beta-4c_{4}\alpha)\frac{a_{f}^{2}}{a^{2}}+4(c_{2}\beta-3c_{3}\alpha)\frac{a_{f}}{a}+(c_{1}\beta-2c_{2}\alpha),
	\end{equation}
	\begin{equation}
	C_{4}=\beta^{2}\mathcal{E}_{\tilde{A}}^{\text{com}}+2M_{\text{pl}}^{2}m^{2}\frac{a^{2}}{\tilde{a}^{2}}\left(12c_{4}\frac{a_{f}^{2}}{a^{2}}+6c_{3}\frac{a_{f}}{a}+c_{2}\right),
	\end{equation}
	\begin{eqnarray}
	C_{5} & = & \alpha\beta\frac{\tilde{a}^{3}\tilde{N}}{Na^{2}a_{f}}\mathcal{E}_{\tilde{\zeta}}^{\text{com}}+3M_{\text{pl}}^{2}\beta^{2}\left(3H^{2}+2\dot{H}\right)\nonumber \\
	&  & +M_{\text{pl}}^{2}m^{2}\bigg[2\left(3c_{3}\frac{N_{f}}{N}+c_{2}\right)\beta^{2}\frac{a_{f}^{2}}{a^{2}}+2\frac{a_{f}}{a}\left(9c_{3}\alpha^{2}-4c_{2}\beta\alpha+2c_{1}\beta^{2}\right)\nonumber \\
	&  & \hspace{4em}+8\frac{a_{f}}{a}\frac{N_{f}}{N}\left(9c_{4}\alpha^{2}-3c_{3}\beta\alpha+c_{2}\beta^{2}\right)+\frac{N_{f}}{N}\left(24c_{3}\alpha^{2}-8c_{2}\beta\alpha+3c_{1}\beta^{2}\right)\nonumber \\
	&  & \hspace{4em}+\left(8c_{2}\alpha^{2}-4c_{1}\beta\alpha+3c_{0}\beta^{2}\right)+\frac{a}{a_{f}}\left(2c_{2}\frac{N_{f}}{N}+c_{1}\right)\alpha^{2}\bigg],
	\end{eqnarray}
	\begin{eqnarray}
	C_{6} & = & \beta\alpha\frac{1}{Na^{3}}\frac{a}{a_{f}}\tilde{a}^{3}\tilde{N}\mathcal{E}_{\tilde{\zeta}}^{\text{com}}+M_{\text{pl}}^{2}m^{2}\frac{\tilde{a}}{a_{f}}\Bigg[2\left((9c_{3}\alpha-2c_{2}\beta)+6\frac{N_{f}}{N}(6c_{4}\alpha-c_{3}\beta)\right)\frac{a_{f}^{2}}{a^{2}}\nonumber \\
	&  & \hspace{4em}+2\left((4c_{2}\alpha-c_{1}\beta)+2\frac{N_{f}}{N}(6c_{3}\alpha-c_{2}\beta)\right)\frac{a_{f}}{a}+\left(2c_{2}\frac{N_{f}}{N}+c_{1}\right)\alpha\Bigg],
	\end{eqnarray}
	\begin{eqnarray}
	C_{7} & = & M_{\text{pl}}^{2}m^{2}\left[18\left(4c_{4}\frac{N_{f}}{N}+c_{3}\right)\frac{a_{f}}{a}+8\left(3c_{3}\frac{N_{f}}{N}+c_{2}\right)+\left(2c_{2}\frac{N_{f}}{N}+c_{1}\right)\frac{a}{a_{f}}\right]\nonumber \\
	&  & +\beta\left(\frac{\tilde{a}^{2}}{aa_{f}}+2\beta\frac{\tilde{a}}{a}\right)\frac{\tilde{N}}{N}\mathcal{E}_{\tilde{\zeta}}^{\text{com}},
	\end{eqnarray}
	\begin{equation}
	C_{8}=\alpha\beta\frac{\tilde{N}}{N}\frac{\tilde{a}}{a_{f}}\mathcal{E}_{\tilde{\zeta}}^{\text{com}}+M_{\text{pl}}^{2}m^{2}\left[\left(2c_{2}\frac{N_{f}}{N}+c_{1}\right)\frac{a}{a_{f}}+2\left(3c_{3}\frac{N_{f}}{N}+c_{2}\right)\right],
	\end{equation}
and
	\begin{equation}
	m_{\tilde{A}\tilde{A}}=-\frac{1}{2}P-2\tilde{X}^{2}P_{,\tilde{X}\tilde{X}}+3\tilde{H}^{2}\lambda_{1}+3\tilde{H}\bar{\phi}'\lambda_{1,\phi}-27\tilde{H}^{2}\tilde{X}\lambda_{2}-75\tilde{H}^{3}\tilde{X}\bar{\phi}'\lambda_{3},
	\end{equation}
	\begin{equation}
	f_{\tilde{A}\phi}=-\bar{\phi}'P_{,\tilde{X}}-2\tilde{X}\bar{\phi}'P_{,\tilde{X}\tilde{X}}+6\tilde{H}\lambda_{1,\phi}-18\tilde{H}^{2}\bar{\phi}'\lambda_{2}-90\tilde{H}^{3}\tilde{X}\lambda_{3},
	\end{equation}
	\begin{equation}
	f_{\tilde{A}\tilde{\zeta}}=2\lambda_{1}+\frac{\bar{\phi}'\lambda_{1,\phi}}{\tilde{H}}-6\tilde{X}\lambda_{2}-15\tilde{H}\tilde{X}\bar{\phi}'\lambda_{3},
	\end{equation}
	\begin{equation}
	w_{\tilde{B}\phi}=-\bar{\phi}'P_{,\tilde{X}}+2\tilde{H}\lambda_{1,\phi}-2\bar{\phi}'\lambda_{1,\phi\phi}-6\tilde{H}^{2}\bar{\phi}'\lambda_{2}+4\tilde{H}\tilde{X}\lambda_{2,\phi}-18\tilde{H}^{3}\tilde{X}\lambda_{3}+6\tilde{H}^{2}\tilde{X}\bar{\phi}'\lambda_{3,\phi},
	\end{equation}
	\begin{eqnarray}
	m_{\phi\phi} & = & \frac{1}{2}P_{,\phi\phi}-\frac{1}{2}P_{,\tilde{X}\phi}\left(3\tilde{H}\bar{\phi}'+\bar{\phi}''\right)-\tilde{X}\left(\bar{\phi}''P_{,\text{XX\ensuremath{\phi}}}+P_{,\text{X\ensuremath{\phi\phi}}}\right)+3\lambda_{1,\phi\phi}\left(2\tilde{H}^{2}+\tilde{H}'\right)\nonumber \\
	&  & -3\tilde{H}\lambda_{2,\phi}\left(3\tilde{H}^{2}\bar{\phi}'+\tilde{H}\bar{\phi}''+2\tilde{H}'\bar{\phi}'\right)-3\tilde{H}^{2}\tilde{X}\lambda_{2,\phi\phi}\nonumber \\
	&  & -6\tilde{H}^{3}\tilde{X}\bar{\phi}'\lambda_{3,\phi\phi}-9\tilde{H}^{2}\lambda_{3,\phi}\left(3\tilde{X}\left(\tilde{H}^{2}+\tilde{H}'\right)+\tilde{H}\bar{\phi}'\bar{\phi}''\right),
	\end{eqnarray}
	\begin{equation}
	w_{\phi\phi}=-\frac{1}{2}P_{,\tilde{X}}-\lambda_{2}\left(3\tilde{H}^{2}+2\tilde{H}'\right)-3\tilde{H}\lambda_{3}\left(2\bar{\phi}'\left(\tilde{H}^{2}+\tilde{H}'\right)+\tilde{H}\bar{\phi}''\right)-2\tilde{H}^{2}\tilde{X}\lambda_{3,\phi},
	\end{equation}
	\begin{eqnarray}
	w_{\tilde{\zeta}\phi} & = & 4\lambda_{1,\phi}-4\lambda_{2}\left(\tilde{H}\bar{\phi}'+\bar{\phi}''\right)-4\tilde{X}\lambda_{2,\phi}-12\lambda_{3}\left(\tilde{X}\left(\tilde{H}^{2}+\tilde{H}'\right)+\tilde{H}\bar{\phi}'\bar{\phi}''\right)\nonumber \\
	&  & +\lambda_{3,\phi}\left(-8\tilde{H}\tilde{X}\bar{\phi}'-12\tilde{X}\bar{\phi}''+6\bar{\phi}'^{2}\bar{\phi}''\right),
	\end{eqnarray}
	\begin{equation}
	g_{\phi\phi}=\frac{1}{2}P_{,\tilde{X}}+\tilde{X}P_{,\tilde{X}\tilde{X}}+9\tilde{H}^{3}\bar{\phi}'\lambda_{3}+3\tilde{H}^{2}\lambda_{2},
	\end{equation}
	\begin{equation}
	g_{\tilde{\zeta}\phi}=6\lambda_{1,\phi}+12\tilde{H}\bar{\phi}'\lambda_{2}+54\tilde{H}^{2}\tilde{X}\lambda_{3}.
	\end{equation}
	
The advantage of using $\{\zeta,\tilde{\zeta},\tilde{E}\}$ is that in (\ref{S2_s_3v}) $\zeta$ has no time derivative explicitly and thus can be further integrated out.
To summarize, the final quadratic action for $\left\{ \tilde{\zeta},\tilde{E}\right\} $
can be written as
\begin{eqnarray}
S_{2}\left[\tilde{\zeta},\tilde{E}\right] & = & \int\mathrm{d}t\frac{\mathrm{d}^{3}k}{\left(2\pi\right)^{3}}\bigg[\mathcal{G}_{11}(\partial_{t}\tilde{\zeta})^{2}+2\mathcal{G}_{12}\partial_{t}\tilde{\zeta}\partial_{t}\tilde{E}+\mathcal{G}_{22}(\partial_{t}\tilde{E})^{2}\nonumber \\
&  & +\Xi\left(\partial_{t}\tilde{\zeta}\,\tilde{E}-\tilde{\zeta}\,\partial_{t}\tilde{E}\right)+\mathcal{W}_{11}\tilde{\zeta}^{2}+2\mathcal{W}_{12}\tilde{\zeta}\tilde{E}+\mathcal{W}_{22}\tilde{E}^{2}\bigg],
\end{eqnarray}
with
\begin{eqnarray}
\mathcal{G}_{11} & = & \frac{4\mathcal{A}_{1}\mathcal{A}_{13}-\mathcal{A}_{4}^{2}}{4\Delta\mathcal{A}_{1}},\label{Gcal11}\\
\mathcal{G}_{12} & = & \frac{2\mathcal{A}_{1}\mathcal{A}_{14}-\mathcal{A}_{4}\mathcal{A}_{5}}{4\Delta\mathcal{A}_{1}},\label{Gcal12}\\
\mathcal{G}_{22} & = & \frac{4\mathcal{A}_{1}\mathcal{A}_{15}-\mathcal{A}_{5}^{2}}{4\Delta\mathcal{A}_{1}},\label{Gcal22}
\end{eqnarray}
and
\begin{equation}
\Xi=\frac{2\mathcal{A}_{1}\mathcal{A}_{11}-\mathcal{A}_{3}\mathcal{A}_{4}}{4\Delta\mathcal{A}_{1}}-\frac{2\mathcal{A}_{1}\mathcal{A}_{9}-\mathcal{A}_{2}\mathcal{A}_{5}}{4\Delta\mathcal{A}_{1}},\label{Xi}
\end{equation}
and
\begin{eqnarray}
\mathcal{W}_{11} & = & -\frac{\mathcal{A}_{2}^{2}-4\mathcal{A}_{1}\mathcal{A}_{6}}{4\Delta\mathcal{A}_{1}}-\frac{\mathrm{d}}{\mathrm{d}t}\left(\frac{2\mathcal{A}_{1}\mathcal{A}_{8}-\mathcal{A}_{2}\mathcal{A}_{4}}{4\Delta\mathcal{A}_{1}}\right),\label{Wcal11}\\
\mathcal{W}_{12} & = & \frac{2\mathcal{A}_{1}\mathcal{A}_{7}-\mathcal{A}_{2}\mathcal{A}_{3}}{4\Delta\mathcal{A}_{1}}-\frac{1}{2}\frac{\mathrm{d}}{\mathrm{d}t}\left(\frac{2\mathcal{A}_{1}\mathcal{A}_{11}-\mathcal{A}_{3}\mathcal{A}_{4}}{4\Delta\mathcal{A}_{1}}+\frac{2\mathcal{A}_{1}\mathcal{A}_{9}-\mathcal{A}_{2}\mathcal{A}_{5}}{4\Delta\mathcal{A}_{1}}\right),\label{Wcal12}\\
\mathcal{W}_{22} & = & \frac{4\mathcal{A}_{1}\mathcal{A}_{10}-\mathcal{A}_{3}^{2}}{4\Delta\mathcal{A}_{1}}-\frac{\mathrm{d}}{\mathrm{d}t}\left(\frac{2\mathcal{A}_{1}\mathcal{A}_{12}-\mathcal{A}_{3}\mathcal{A}_{5}}{4\Delta\mathcal{A}_{1}}\right).\label{Wcal22}
\end{eqnarray}
In the above, $\Delta$ is given in (\ref{Del_def}), $\mathcal{A}_1,\cdots,\mathcal{A}_{15}$ are given in (\ref{Acal1})-(\ref{Acal15}).

% % % % % % % % % % % % % %
%%%%%%%%%%%%%%%%%%

\providecommand{\href}[2]{#2}\begingroup\raggedright\endgroup

\end{document}